\documentclass[12pt,thmsa,arabtex]{article}
\usepackage{amsfonts}
\usepackage{graphicx}

\input{tcilatex}

\begin{document}

\begin{center}
{\Large Quantitative aspects of entanglement in the optically
driven quantum dots}

{\Large \ }

\textbf{A. S.-F. Obada$^{1}$ and M.
Abdel-Aty$^{2,}$}\footnote[3]{E-mail:
abdelatyquantum@gmail.com\newline On leave from Mathematics
Department, Faculty of Science, Sohag University, 82524 Sohag,
Egypt}

{\small $^1$Mathematics Department, Faculty of Science, Al-Azhar
University,
Nasr City, cairo, Egypt\\[0pt]
$^{2}$Mathematics Department, College of Science, Bahrain
University, 32038 Kingdom of Bahrain }

\bigskip
\end{center}

We present a novel approach to look for the existence of maximum
entanglement in a system of two identical quantum dots coupled by
the F\"{o} rster process and interacting with a classical laser
field. Our approach is not only able to explain the existing
treatments, but also provides further detailed insights into the
coupled dynamics of quantum dots systems. The result demonstrates
that there are two ways for generating maximum entangled states, one
associated with far off-resonance interaction, and the other
associated with the weak field limit. Moreover, it is shown that
exciton decoherence results in the decay of entanglement.\bigskip

\section{Introduction}

Quantum dots are semiconductor structures containing a small
number of electrons within a region of space with typical sizes in
the sub-micrometer range \cite{bra05}. Coupling of two quantum
dots leads to double quantum dots, which in analogy with atomic
and molecular physics, is described as two-level systems with
controllable level-spacing and one additional transport electron
\cite{pas06}. This rather suggests the analogy with a simple model
for an atom, in particular if it comes to interaction with
external fields such as photons or phonons. Many properties of
such systems can be investigated by transport, if the dots are
fabricated between contacts acting as source and drain for
electrons which can enter or leave the dot. The possibility of
using pairs of quantum dots coupled by the dipole-dipole
interaction as effective three- or four-level systems whose
transmission for an optical beam at some frequency may be switched
on or off using a second optical beam has been explored
\cite{gea06}. In contrast to real atoms, quantum dots are open
systems with respect to the number of electrons which can easily
be tuned with external parameters such as gate voltages or
magnetic fields \cite{ste06,los98}.

The experimental realization of optically induced entanglement of
excitons in a single quantum dot \cite{che00} and theoretical
study on coupled quantum dots \cite{rei00} have been reported
recently. In those investigations a classical laser field is
applied to create the electron-hole pair in the dot(s). Several
groups have performed transport experiments with double quantum
dots, with lateral structures offering experimental advantages
over vertical dots with respect to their tunability of parameters
\cite{wie03}. However, in contrast to atomic systems carrier
lifetimes in the solid state are much shorter because of the
continuous density-of-states of charge excitations and stronger
environment coupling \cite{he06,pot97}.

Recently a major advancement in the field has come from different
types of local optical experiments, that allow the investigation
of individual quantum dots thus avoiding inhomogeneous broadening
and simple coherent-carrier control in single dots \cite{hoh99}.
Some quantum information processing schemes have been proposed
exploiting exchange and/or direct Coulomb interactions between
spatially separated excitonic qubits in coupled quantum dots
systems \cite{rei00,bio00,lov03,yin06}. Using far-field light
excitation to globally address two and three quantum dots in a
spatially symmetric arrangement and preparations of both Bell and
Greenberger-Horne-Zeilinger entangled states of excitons have been
discussed \cite{bio00,lov03}. Also, an alternative scheme for a
three-qubit entangled state generation by nonlinear optical state
truncation has been introduced \cite{sai06}.

Based on adiabatic elimination treatment \cite{pas06}, the
creation of two-particle entangled states in a system of coupled
quantum dots have been discussed. The authors of this study
avoided dealing with the exact solution of the problem and
employed the adiabatic elimination as an approximate treatment.
However, one should be aware that its predictions for any coupled
system need to be checked against the exact solution of the
complete coupled equations. In order to avoid such limitations,
one must begin by ignoring any approximation and try to find an
analytical solution of the coupled equations that govern the
system. At this point, there is no generally established approach
that can provide a complete description of the dynamics for the
system. It is the purpose of this paper to present such an
approach with illustrative applications. The basic idea relies on
the discretization of the coupled system, which is thus replaced
in the formulation by only linear solvable equation. Related
treatments based on either adiabatic elimination
\cite{pas06,naz04,pas03}, discussing entangled state generation
conditions, or the coupled equations without the detuning
dependence, have been presented in the literature \cite{los98}.

What we study and present below is essentially a most general case
of the complete equations of the two-quantum dots system. Most
interestingly, it is shown that features of the degree of
entanglement are influenced significantly by different values of
the involved parameters and exciton decoherence. With this
approach we could create a two-particle entangled state between
the vacuum and bi-exciton states or single-exciton entangled
state, without using the approximation method adopted in previous
studies.

The outline of this paper is arranged as follows: in section 2, we
give notation and definitions of the model. To reach our goal, an
analytical approach for obtaining exact-time dependent expressions
for the probability amplitudes is developed in section 2.1 and
exciton decoherence is discussed in section 2.2. Having obtained
the solution, in section 3, we analyze the time evolution of the
populations of the quantum levels for various values of the system
parameters. In section 4, we study the evolution of the degree of
entanglement, measured by the negativity measure for the partial
transpose density matrix. Finally, our conclusion\ is presented in
section 5%
\textbf{.}

\section{Model}

The model under consideration consists of two identical quantum
dots coupled by the F\"{o}rster process
\cite{nie00,ola06,kan98,vri00}. This process originates from the
Coulomb interaction whereby an exciton can hop between the two
dots \cite{nie00}. The quantum dots contain no net charge and
interact with a high frequency laser pulse. This in effect means
that the present model has three processes: (i) the coupling of
the carrier system with a classical laser field, (ii) the interdot
F\"{o}rster interaction and (iii) the single-exciton, keeping in
mind the fact that all constant energy terms may be ignored. The
total Hamiltonian for the quantum-dot system is given by
\cite{rod01},
\begin{eqnarray}
\hat{H} &=&\frac{\hbar }{2}\sum\limits_{j=1,2}\left( \Omega \exp
\left[
-i\omega t+i\phi \right] \widehat{e}_{j}^{\dagger }\widehat{\psi }%
_{j}^{\dagger }+\Omega ^{\ast }\exp \left[ i\omega t-i\phi \right]
\widehat{%
\psi }_{j}\widehat{e}_{j}\right)  \nonumber \\
&&-\frac{1}{2}\sum\limits_{j=1}^{2}\left\{
\sum\limits_{k=1}^{2}\hbar \eta \left( \widehat{e}_{j}^{\dagger
}\widehat{\psi
}_{k}\widehat{e}_{k}\widehat{%
\psi }_{j}^{\dagger }+\widehat{\psi }_{j}\widehat{e}_{k}^{\dagger
}\widehat{%
\psi }_{k}^{\dagger }\widehat{e}_{j}\right) -\varepsilon \left(
\widehat{e}%
_{j}^{\dagger }\widehat{e}_{j}-\widehat{\psi }_{j}\widehat{\psi }%
_{j}^{\dagger }\right) \right\} ,  \label{ham2}
\end{eqnarray}%
where $\Omega $ represents the laser-quantum dot coupling and
$\hbar \Omega =\mu E$, where $\mu $ is the coupling strength and
$E$ is the laser field amplitude. The parameters $\omega $ and
$\phi $ describe the angular frequency and phase of the laser
field, respectively. The operator
$\widehat{%
e}_{j}(\widehat{\psi }_{j})$ is the electron (hole) annihilation
operator and $\widehat{e}_{j}^{\dagger }(\widehat{\psi
}_{j}^{\dagger })$ is the electron (hole) creation operator in the
$j^{\underline{th}}$ quantum dot. We denote by $\varepsilon $ the
band gap energy of the quantum dot and $\eta $ the inter-dot
process hopping rate.

For a coupled two-quantum-dot system, it is useful to write the
Hamiltonian of equation (\ref{ham2}) in the basis $|0\rangle
=|0,0\rangle ,$ $|1\rangle =\left( |1,0\rangle +|0,1\rangle
\right) /\sqrt{2},$ and $|2\rangle =|1,1\rangle $ which describe
the vacuum state, the single exciton state, and the biexciton
state, respectively. In this system the antisymmetric state
$|a\rangle =\left( |1,0\rangle -|0,1\rangle \right) /\sqrt{2}$ is
completely decoupled from the remaining states. Then the simple
three-state representation of the two-quantum-dot system can be
employed with $|0\rangle ,$ $|1\rangle $ and $|2\rangle .$ Since
the density matrix of the system is diagonal and the symmetric
state $|1\rangle $ is a maximally entangled state, an entanglement
can be produced in this model by a suitable population of the
state $|1\rangle$.

Applying the rotating wave approximation and a unitary
transformation, the resulting Hamiltonian may be written as

\begin{equation}
H=2\hbar \Delta \widehat{A}_{22}+\hbar (\Delta -\eta
)\widehat{A}_{11}+\frac{%
\hbar }{\sqrt{2}}\left[ \frac{{}}{{}}\Omega e^{i\phi }\left(
\widehat{A}%
_{01}+\widehat{A}_{12}\right) +H.c.\right] ,  \label{h2}
\end{equation}%
where $\widehat{A}_{ij}=|i\rangle \langle j|,$ related to the
above
states $%
|0\rangle ,$ $|1\rangle $ and $|2\rangle .$ We denote by $\Delta $
the detuning of the laser frequency from exact resonance ($\hbar
\Delta =\varepsilon -\hbar \omega )$.

It is worth mentioning here that, one can take advantage of the
F\"{o}rster interaction between two quantum dots and apply a
finite rectangular pulse and sub-picosecond duration to generate a
Bell state such as $\alpha |00\rangle +\beta |11\rangle $ where
$|11\rangle $ denotes the simultaneous presence of two excitons in
a double dot structure. Also, formations of an entangled state
between the vacuum and the exciton (or the biexciton) state have
been discussed \cite{kis04,zha03}.

\subsection{An analytic solution}

We devote the present section to find an explicit expression for
the wave function in Schr\"{o}dinger picture. We use an analytic
approach that seeks to reduce the coupled equations system
(probability amplitudes) to a solvable linear equation in order to
study in detail the types of interaction that exist between them.
To reach our goal we assume that the wave function of the complete
system may be expanded in terms of the well known eigenstates
$\left\vert i\right\rangle $, $(i=0,1,2)$, namely
\begin{equation}
\left\vert \Psi (t)\right\rangle =B_{0}(t)\left\vert
0\right\rangle +B_{1}(t)\left\vert 1\right\rangle
+B_{2}(t)\left\vert 2\right\rangle . \label{weq}
\end{equation}%
The time dependence of the amplitudes in equation (\ref{weq}) is
governed by the Schr\"{o}dinger equation with the Hamiltonian
given by equation
(\ref{h2}%
), therefore we obtain
\begin{equation}
i\frac{\partial B_{j}(t)}{\partial t}=\sum\limits_{k=0}^{2}\xi
_{jk}B_{k}(t), \label{Pro}
\end{equation}%
where $\xi _{jk}=\left\langle j\right\vert \hat{H}\left\vert
k\right\rangle . $ In this case and using equations (\ref{h2}) and
(\ref{weq}), we
obtain $%
\xi _{01}=\left( \xi _{10}\right) ^{\ast }$ $=\xi _{12}=\left( \xi
_{21}\right) ^{\ast }$ $=\Omega ^{\prime }e^{-i\phi },$ $\xi
_{11}=\Delta -\eta ,$ $\xi _{22}=2\Delta ,$ $\Omega ^{\prime
}=\Omega /\sqrt{2}$ \ otherwise $\xi _{ij}=0.$

In order to solve equation (\ref{Pro}), we introduce the following
function \cite{abd87}
\begin{equation}
G(t)=B_{0}(t)+xB_{1}(t)+yB_{2}(t),
\end{equation}%
which leads to the equation
\begin{equation}
i\frac{dG(t)}{dt}=\beta \left( B_{0}(t)+\frac{\gamma _{1}}{\beta
}B_{1}(t)+%
\frac{\gamma _{2}}{\beta }B_{2}(t)\right) ,
\end{equation}%
where $\beta =x\xi _{10},\gamma _{1}=\xi _{12}+x\xi _{11}+y\xi
_{21}$ and${\ \ }\gamma _{2}=x\xi _{12}+y\xi _{22}.$ Now let us
seek a solution of $G(t)$ such that $\dot{G}(t)=-izG(t)$. This
holds if and only if $z=\beta ,x=\gamma _{1}/\beta $ and $\
y=\gamma _{2}/\beta .$

After some minor algebra this leads to a qubic equation which
contains $3$ eigenvalues (to determine the $z_{i}$) corresponding
to the same number of the eigenfunctions $G_{j}(t)=G_{j}(0)\exp
(-iz_{j}t).$ Using equation (5), one can write
\begin{equation}
G_{j}(t)=\sum\limits_{l=1}^{3}O_{jl}\overline{B}_{l}(t),
\end{equation}%
where $O_{jl}$ is a $3\times 3$ matrix whose elements are $%
O_{j1}=1,O_{j2}=x_{j}$ and $O_{j3}=y_{j}$ and
$\overline{B}_{l}=B_{l+1}.$

Now, we can express the unperturbed state amplitude
$\overline{B}_{j}(t)$ in terms of the dressed state amplitudes
$G_{j}(t)$ in this form
$\overline{B}%
_{i}(t)=\sum\limits_{j=1}^{3}\left( O^{-1}\right)
_{ij}G_{j}(0)\exp (-iz_{j}t).$ \ Using the above equations, we
have
\begin{equation}
\overline{B}_{j}(t)=\lambda _{j1}\exp \left( -iz_{1}t\right)
+\lambda _{j2}\exp \left( -iz_{2}t\right) +\lambda _{j3}\exp
\left( -iz_{3}t\right) , \label{sol}
\end{equation}%
where

\begin{eqnarray}
z_{j} &=&-\frac{\xi _{10}}{3}\left[ \alpha _{1}+2\left( \alpha
_{1}^{2}-3\alpha _{2}\right) ^{\frac{1}{2}}\cos \left(
\frac{1}{3}\arccos
\left( \frac{9\alpha _{1}\alpha _{2}-2\alpha _{1}^{3}-27\alpha _{3}}{%
2(\alpha _{1}^{2}-3\alpha _{2})^{3/2}}\right) +[j-1]/3\right)
\right] ,
\nonumber \\
\alpha _{1} &=&-\frac{\xi _{11}+\xi _{22}}{\xi _{01}^{\ast }},{\ \
\ \
}%
\alpha _{2}=-\frac{\xi _{11}\xi _{22}+2|\xi _{01}|^{2}}{\left( \xi
_{01}^{\ast }\right) ^{2}},{\ \ \ }\alpha _{3}=-\frac{\xi
_{22}}{\xi _{01}^{\ast }}.  \nonumber
\end{eqnarray}%
The parameter $\lambda _{ij}=\left( O^{-1}\right) _{ij}G_{j}(0)$
where
$%
\left( O^{-1}\right) _{ij}$ is the ij element of the matrix
$O^{-1}$ which is the inverse of the matrix $O$. We have thus
completely determined an analytic solution of the coupled quantum
dots system in presence of the detuning parameter and phase.

\subsection{Quantum decoherence}

The original meaning of decoherence was specifically designated to
describe the loss of coherence in the off-diagonal elements of the
density operator in the energy eigenbasis \cite{moy06}. Amongst
the most crucial requirements for the implementation of quantum
logic devices is a high degree of quantum coherence. Coherence is
lost when a qubit interacts with other quantum degrees of freedom
in its environment and becomes entangled with them. Exciton
decoherence in semiconductor quantum dots is affected by many
environmental effects, however it is dominated by acoustic phonon
scattering at low temperatures \cite{tak99}. The decoherence
effects due to the exciton--acoustic-phonon coupling on the
generation of an exciton maximally entangled state in quantum dots
were studied in \cite{yi00}. This process is governed by the
Hamiltonian \cite{qui99}

\begin{equation}
\widehat{H}_{T}=\widehat{H}+\sum_{k}\omega _{k}a_{k}^{\dag
}a_{k}+\sum_{k}g_{k}J_{z}\left( a_{k}^{\dag }+a_{k}\right) ,
\end{equation}%
where $\widehat{H}$ is given by equation (1) and $a_{k}^{\dag
}(a_{k})$ stands for the creation (annihilation) operator of the
acoustic phonon with wave vector $k$ and $g_{k}$ the coupling
between the dots and the field. By the general procedure, we can
deduce a master equation for the density
operator $\rho (t)$ of the total system in the following form%
\begin{equation}
\frac{\partial }{\partial t}\rho (t)=J\rho (t)+\pounds _{1}\rho
(t), \label{dec}
\end{equation}%
with
\[
\pounds _{1}\rho (t)=-i\Gamma \lbrack J_{z},[J_{z},\rho (t)]],
\]%
where the superoperator $J$ is defined as $J\rho
(t)=-i[\widehat{H},\rho (t)] $ and $\Gamma $ is the decoherence
rate [5]. However its dependence on the mode distribution of
phonons as well as a cutoff frequency is given by
\cite%
{rod00},%
\begin{equation}
\Gamma =\int d\omega ^{\prime }\omega ^{\prime n}\exp \left(
\frac{-\omega ^{\prime }}{\omega _{c}}\right) (1+2N(\omega
^{\prime },T)),
\end{equation}%
with $n$ depending on the dimensionality of the phonon field,
$\omega _{c}$ is a cut-off frequency and $N(\omega ^{\prime },T)$
is the phonon occupation factor. Here, we consider pure
decoherence effects that do not involve energy relaxation of
excitons. The solution of equation (\ref{dec}) can be formally
written as
\begin{equation}
\rho (t)=\exp [t\left( J+\pounds _{1}\right) ]\rho (0).
\end{equation}%
Here $\rho (0)$ is the initial state of the system. The
decoherence parameter $\Gamma $\ is temperature dependent and it
amounts for
$20-60$ $%
\mu eV$ for typical semiconductor quantum dots in a temperature
range
from $%
10$ $K$ to $30$ $K$ \cite{tak99,rod00}. The results of this
analysis are more closely related to experimental situations,
which are usually strongly affected by decoherence and relaxation.
However, because time scales are very long the relaxation
processes are not considered here.

\section{Occupation probabilities}

\begin{figure}[tbph]
\begin{center}
\includegraphics[width=6cm]{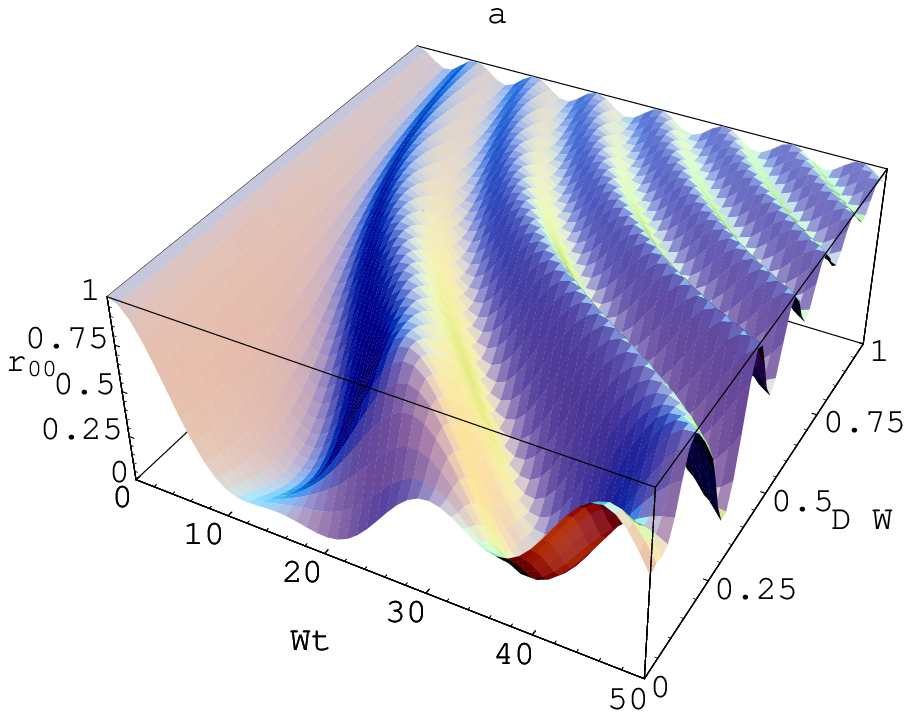}
\includegraphics[width=6cm]{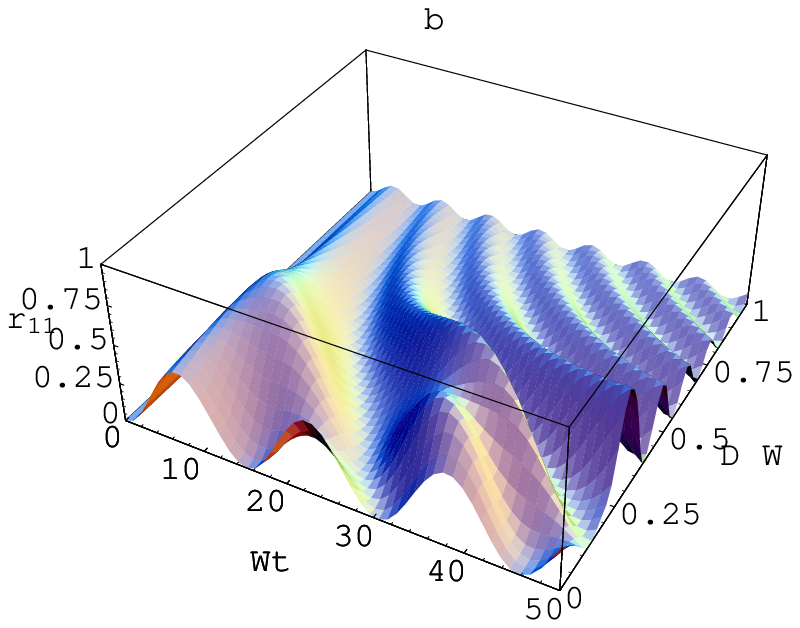}
\includegraphics[width=6cm]{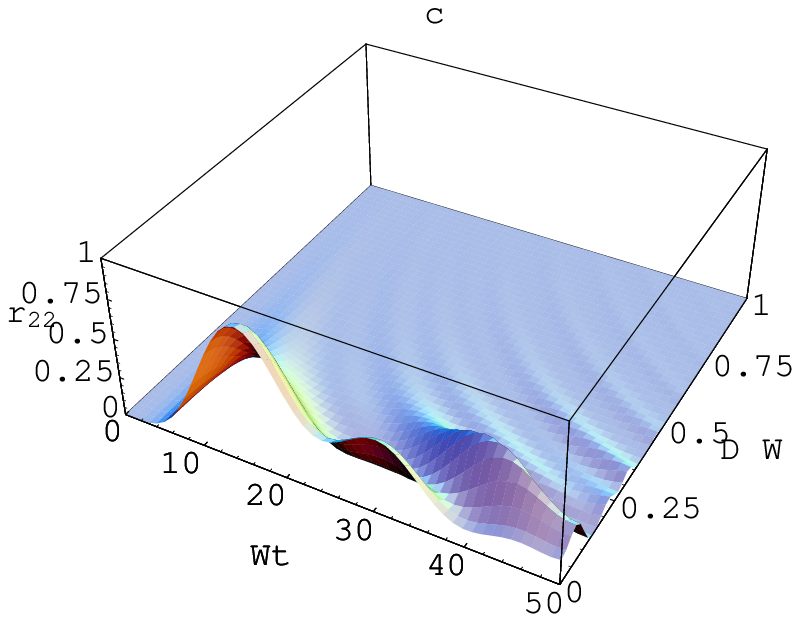}
\end{center}
\caption{The probability amplitudes as functions of $\Omega t$ and
$\Delta /\Omega $. The parameters used in these figures are
$\protect\phi
=0,\protect%
\eta /\Omega =0.1$ and $\Gamma =0$.} \label{fig1}
\end{figure}
By making use of the theoretical treatment in the previous
section, one can investigate the statistical properties of the
system. Using the final
state $%
\left\vert \Psi (t)\right\rangle $ or $\rho (t)$ all relevant
quantities can be computed. In this section, our motivation is to
investigate the occupation probabilities associated with two
identical quantum dots coupled by the F\"{o}rster process and
interacting with a classical laser field. The expressions $\rho
_{ii}(t)=|B_{i}(t)|^{2},$ ($i=0,1,2$) represent the probabilities
that at time $t,$ the coupled quantum dots is in the
state $%
|i\rangle $.

In figure 1, we plot the probability amplitudes as functions of
the dimensionless parameters $\Omega t$ and $\Delta /\Omega $. The
parameters used in these figures are $\phi =0,\eta /\Omega =0.1$
and $\Gamma =0$. It implies that the complete Rabi oscillations
between the localized states occur. Once the detuning parameter is
taken into account, the populations of biexciton state is
decreased (see figure 1c). On further increase of the detuning
parameter $\Delta $ one finds that the occupation probability of
this level tends to zero, while the oscillations of the other
levels show fast oscillations with small amplitudes (see figures
1). From this point of view, the transition can be considered as
existing only between two
states $%
|0>$ and $|1>$for larger detuning, which means that the biexciton
state is decoupled.

\begin{figure}[tbph]
\begin{center}
\includegraphics[width=6cm]{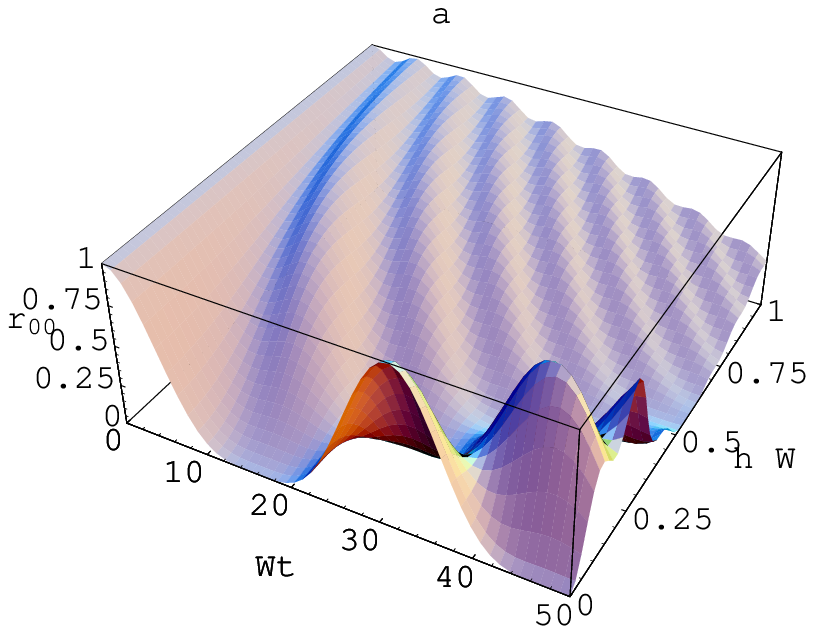}
\includegraphics[width=6cm]{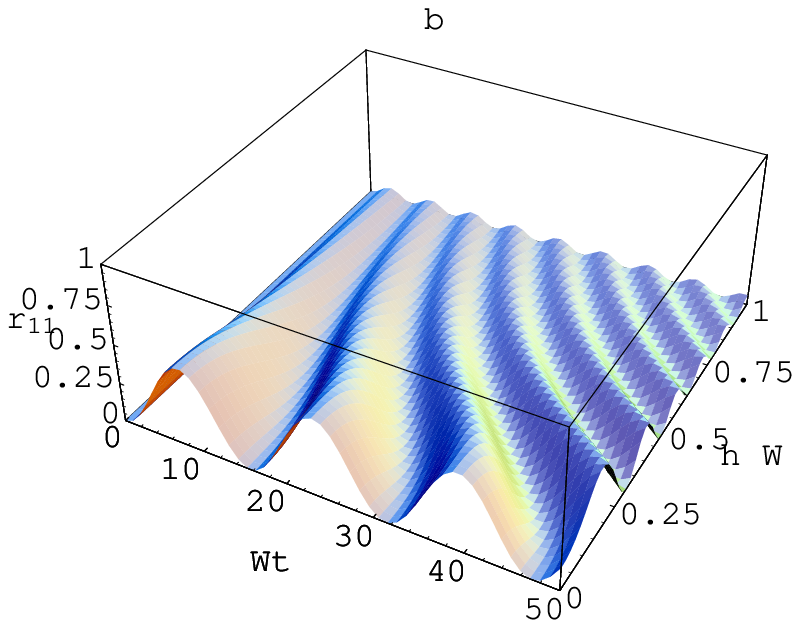}
\includegraphics[width=6cm]{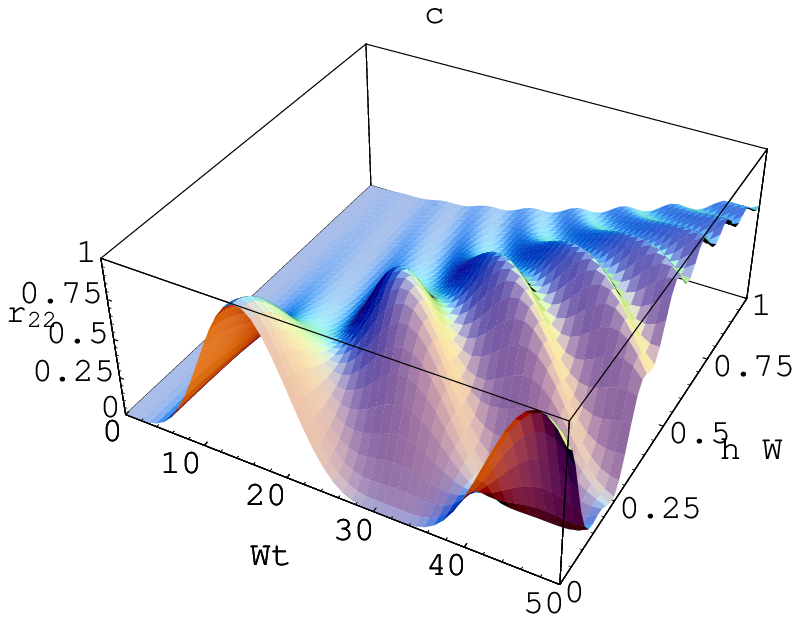}
\end{center}
\caption{The probability amplitudes as functions of $\Omega t$ and
$\protect%
\eta /\Omega $. Calculations assume that $\protect\phi =0,\Delta
/\Omega =0,$ and $\Gamma =0$.} \label{fig2}
\end{figure}
One, possibly not very surprising, principal observation is that
the numerical calculations corresponding to the same parameters,
which have been considered in \cite{pas06}, gives nearly the same
behavior, but with different scaled time. The important
consequence of this observation is that using smaller values of
the laser-quantum dot coupling, the creation of a two-particle
entangled state between the vacuum and the bi-exciton states or a
two-particle entangled state between vacuum and single-exciton
states, depends essentially on controlling the detuning parameter.

Apart from helping us to know different conditions for creating
maximum entangled states, using our exact solution, that would
otherwise be very hard to figure out, some of our testing results
are of interest. A particularly non-intuitive one is the
following: previously, we used to think that increasing the
detuning parameter always decreases the populations of the
single-exciton state or bi-exciton state, depending on its
relation with the laser-dot coupling parameter. But, in figure 2
even with fixing the detuning parameter to be $\Delta =0,$ a
maximum entangled state between the vacuum and bi-exciton states
is obtained when $\eta /\Omega $ takes larger values (see figure
2). \ When we first encountered this result (quite rare depending
on the detuning), the problem becomes interesting and needs more
investigations. Promisingly, we find that further small reductions
in the laser-dot coupling parameter contribution will lead to
significant improvements in entanglement.

Whereas the previous section dealt with the general behavior of
the probability amplitudes and their indications to the entangled
states generation, the next section introduces another view of the
realization of the maximum entanglement.

\section{ Degree of entanglement}

The characterization and classification of entanglement in quantum
mechanics is one of the cornerstones of the emerging field of
quantum information theory. Although an entangled two-qubit state
is not equal to the product of the two single-qubit states
contained in it, it may very well be a convex sum of such
products. In general it is known that microscopic entangled states
are found to be very stable, for example electron-sharing in
atomic bonding and two-qubit entangled photon states generated by
parametric down conversion [19].

In this article, we take the measure of negative eigenvalues for
the partial transposition of the density operator as an
entanglement measure. According to the Peres and Horodecki's
condition for separability \cite{per96,hor96}, a two-qubit state
for the given set of parameter values is entangled if and only if
its partial transpose is negative. The measure of entanglement can
be defined in the following form \cite{lee00,lee00a}
\begin{equation}
E_{\rho }\left( t\right) =\max \left( 0,-2\sum\limits_{i}\lambda
_{i}\right) ,
\end{equation}%
where the sum is taken over the negative eigenvalues of the
partial transposition of the density matrix $\rho $ of the
system$.$ In the two qubit system ($C^{2}\otimes C^{2}),$ it can
be shown that the partial transpose of the density matrix can have
at most one negative eigenvalue \cite{hor96}\textrm{.}

The entanglement measure then ensures the scale between $0$ and
$1$ and monotonously increases as entanglement grows. An important
situation is that, when $E_{\rho }\left( t\right) =0$ the two
qubits are separable
and $%
E_{\rho }\left( t\right) =1$ indicates maximum entanglement
between the two qubits. It was proved \cite{mes03} that the
negativity is an entanglement monotone, and hence is a good
entanglement measure.

In figure 3, we plot the entanglement degree $E_{\rho }\left(
t\right) $ as a function of the dimensionless parameter $\Omega t$
and $\Delta
/\Omega $ $%
(\eta /\Omega )$. From figure 3a, we see that the first maximum
entanglement as well as the disentanglement ($E_{\rho }\left(
t\right) =0$) occurs at earlier times when the detuning parameter
is increased. These results are thus not in conflict with the
well-established theory of adiabatic elimination. As soon as we
take the detuning effects into consideration it is easy to realize
the decreasing of the amplitudes of the oscillations with
increasing the value of the detuning parameter. Furthermore if we
take the parameter $\Delta $ large enough then one can see that
the entanglement degree tends to zero and the quantum dots become
disentangled. This means that, any change of the detuning
parameter leads to changing in the entanglement. It is interesting
also seeing that the number of oscillations is increased with
increasing the detuning however, with smaller amplitude. In all
these cases, it should be noted that the entanglement vanish for
some periods of the interaction time (except for the case $\Delta
=\eta)$.
\begin{figure}[tbph]
\begin{center}
\includegraphics[width=7cm]{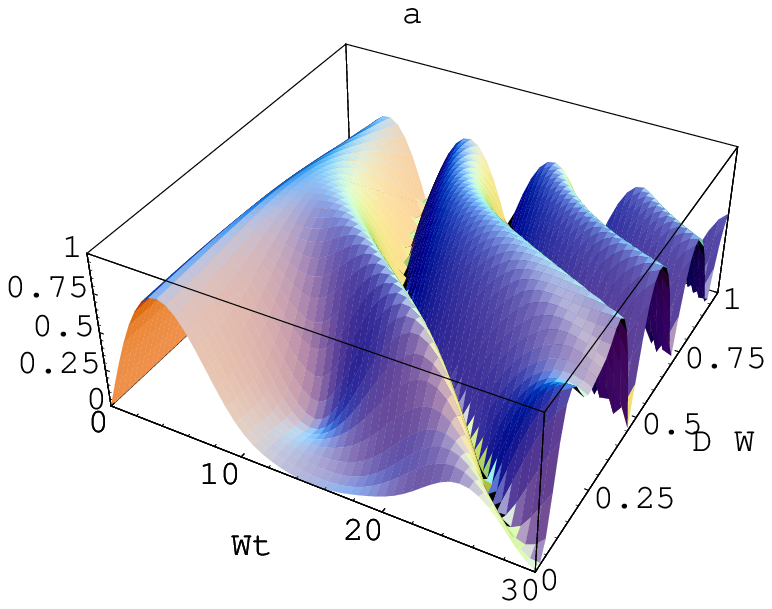}
\includegraphics[width=7cm]{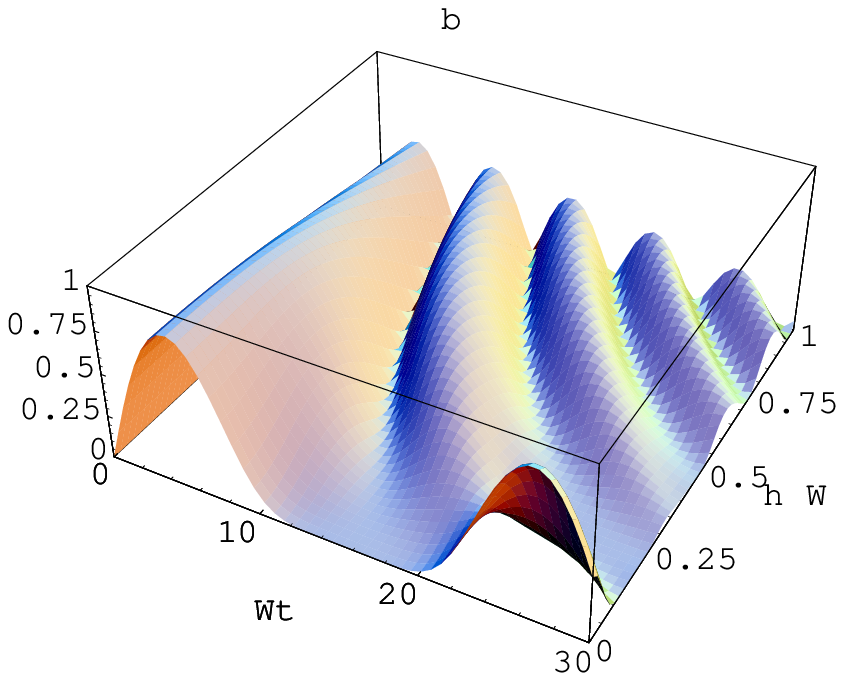}
\end{center}
\caption{The entanglement $E_{\protect\rho }\left( t\right) $ as a
function of $\Omega t$ and $\Delta /\Omega $ ($\protect\eta
/\Omega )$. The parameters used in these figures are (a)
$\protect\phi =0,\protect\eta
/\Omega =0.1$ and $\Gamma =0,$ (b) $\Delta /\Omega =0$, and
$\Gamma =0.$}
\end{figure}

For a very small value of $\eta $ (say $\eta =0.01)$, the
situation becomes surprisingly interesting, at the initial period
of the interaction time the entanglement is strong between the two
quantum dots, but as the time goes on we have seen long survival
of the disentanglement. This result indicates that the quantum
dots will return to a pure state and completely disentangle from
each other for a long period of the interaction time ($12\leq
\Omega t\leq 19$). Finally, we may say that, it is possible to
obtain a long surviving disentanglement using small values of
inter-dot process hopping rate. Which means that the inter-dot
process hopping rate plays an important role in the quantum
entanglement.

\textrm{An interesting question is whether or not the entanglement
is affected by different values of the decoherence parameter
}$\Gamma
/\Omega $%
\textrm{. }Figure 4a displays the effect of $\Gamma /\Omega $ on
the entanglement, where $\Gamma /\Omega =0.01$. Evidently the
evolution dynamics of $E_{\rho }\left( t\right) $ is sensitive to
changes in the decoherence parameter $\Gamma /\Omega $. In the
long time limit, the two quantum dots will be damped into their
vacuum state due to the decoherence effect and entanglement decays
to zero. Which means that the decoherence plays an important role
in the reduction of the degree of entanglement. If the decoherence
parameter is increased further, for the system with fixed values
of the inter-dot coupling, the decrease in the amount of
entanglement is faster (see figure 4b). It is likely that future
source improvements will give values close to those expected for
different initial states: the laser-dot coupling must be reduced
to obtain sufficient entanglement to generate maximum entangled
states. The oscillations in degree of entanglement between the
quantum dots quickly damp out with an increase
in $%
\Gamma /\Omega $. The subsystems will disentangle from each other
and the steady state is reached at earlier interaction time. The
change in $(\eta /\Omega )$ does not show much effect in the
general structure of the degree of entanglement in contrast to the
case\ $\Gamma =0$

\begin{figure}[tbph]
\begin{center}
\includegraphics[width=7cm]{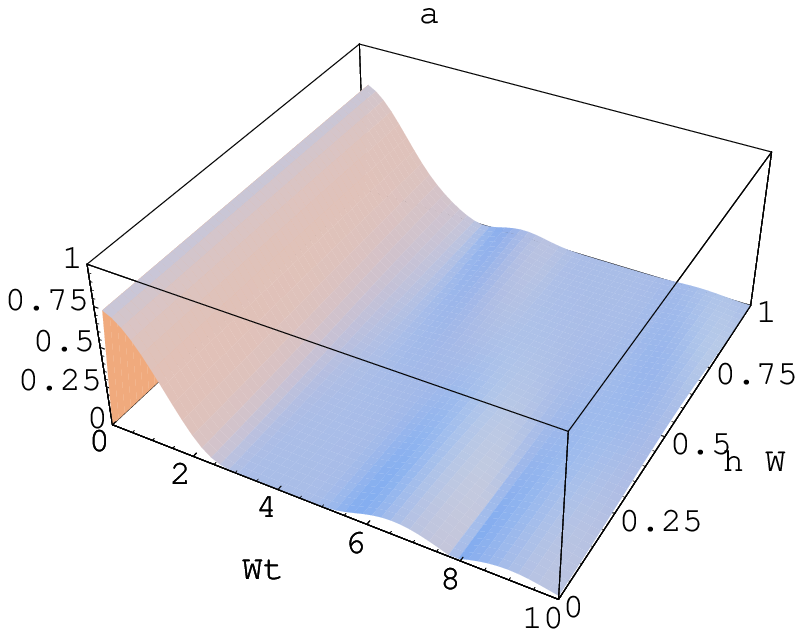}
\includegraphics[width=7cm]{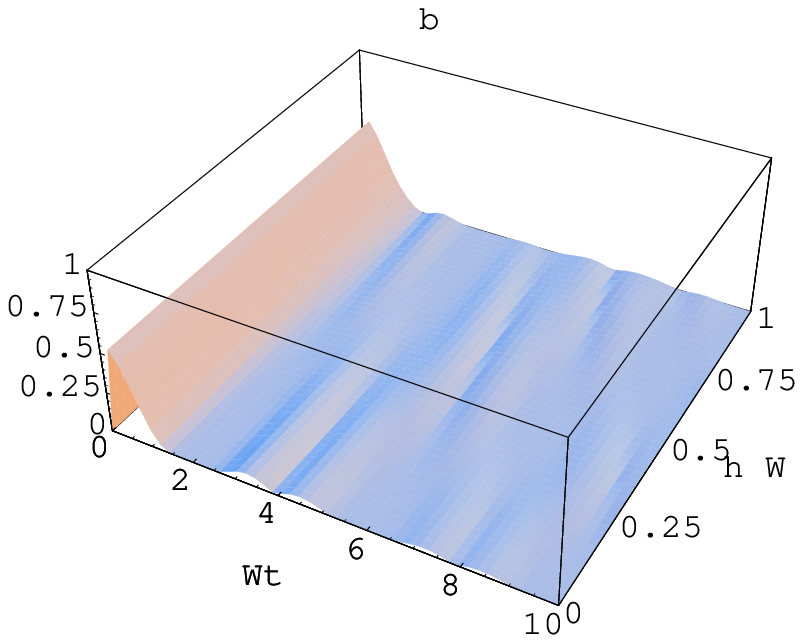}
\end{center}
\caption{The entanglement $E_{\protect\rho }\left( t\right) $ as a
function of $\Omega t$ and $\protect\eta /\Omega $. The parameters
used in these figures are (a) $\Delta /\Omega =0$, and $\Gamma
/\Omega =0.01$ and (b)
$%
\Delta /\Omega =0$ and $\Gamma /\Omega =0.05.$}
\end{figure}

\textrm{\ }It would be also worthwhile to use different initial
states setting, which would strongly help in creating the maximum
entangled states. In a recent experimental work \cite{fas06}, it
has been demonstrated that quantum superpositions and entanglement
can be surprisingly robust. \ This adds to the growing
experimental evidence that robust manipulation of entanglement is
feasible \cite{the04} with today's technology. Entangling many
degrees of freedom, or equivalently many qubits (quantum dots),
remains a challenge, however, these experimental results are
encouraging
\cite{fas06}%
. Also, i\textrm{n connection to the foundations of quantum
theory, a deeper understanding of entanglement decoherence is
expected to lead to new insights into the foundations of quantum
mechanics} \textrm{\ \cite{gis96}.}

\textrm{The remaining task is to identify and compare the results
presented above for the entanglement degree with another accepted
entanglement measure such as the concurrence \cite{min05}. One,
possibly not very surprising, principal observation is that the
numerical calculations corresponding to the same parameters, which
have been considered above, give nearly the same behavior. This
means that estimating the entanglement either using the negativity
or concurrence measures gives qualitatively the same results. }

\section{Conclusion}

We presented an analytical treatment for performing a maximum
entangled states between two qubits in adjacent semiconductor
quantum dots, formed through the interdot F\"{o}rster interaction.
The developed approach is capable of providing exact solutions to
a class of problems which have only been treated approximately
through previous studies. An important aspect is the insight
gained by the possibility to combine the exact solution with the
numerical treatments to generate maximum entangled states. More
explicitly, in the exciton system, the large values of the
detuning helps in generating maximum entangled states. \
Nevertheless, the calculations indicate the maximum entangled
states can still exist, even for the resonant case, when the
electron and hole are driven by a suitable laser field (weak field
limit).

We have extended our studies by giving an analysis and explanation
of the predicted entanglement taking into account the decoherence
effect. A remarkable property of the decoherence effect is that
entanglement can fall abruptly to zero for a very long time and
the entanglement will not be recovered i.e. the state will stay in
the disentanglement separable state. \textrm{Needless to say,
there is still much work to be done and technical and general
questions to be addressed: in particular, }sources of three and
four entangled quanta have very recently been reported
\cite{wal04}, which in turn allows one to extend the question
beyond two quantum dots to many-particle systems. This is also an
exciting area for future study, both theoretically and
experimentally.

\bigskip

\textbf{Acknowledgments}: We are grateful to the referees for very
constructive comments and for suggesting various improvements in the
manuscript. Also, we would like to thank E. Paspalakis for
stimulating communications.

\end{document}